\documentclass[]{spie}  
\usepackage[dvips]{graphicx}

\title{Wide-Field Surveys from the SNAP Mission} 

\author{\hspace{-6pt}\parbox{17cm}{\center A.~Kim\supit{a}, C.~Akerlof\supit{b}, G.~Aldering\supit{a},
R.~Amanullah\supit{c},
P.~Astier\supit{d}, E.~Barrelet\supit{d},
C.~Bebek\supit{a}, L.~Bergstr\"{o}m\supit{c},
J.~Bercovitz\supit{a}, G.~Bernstein\supit{e}, M.~Bester\supit{f},
A.~Bonissent\supit{g}, C.~Bower\supit{h},
W.~Carithers\supit{a},
E.~Commins\supit{f}, C.~Day\supit{a},
S.~Deustua\supit{i}, R.~DiGennaro\supit{a}, A.~Ealet\supit{g},
R.~Ellis\supit{j},
M.~Eriksson\supit{c}, A.~Fruchter\supit{k},
J-F.~Genat\supit{d}, G.~Goldhaber\supit{f},
A.~Goobar\supit{c}, D.~Groom\supit{a},
 S.~Harris\supit{f},
P.~Harvey\supit{f}, H.~Heetderks\supit{f}, S.~Holland\supit{a},
D.~Huterer\supit{l}, A.~Karcher\supit{a},
W.~Kolbe\supit{a}, B.~Krieger\supit{a}, R.~Lafever\supit{a},
J.~Lamoureux\supit{a},
M.~Lampton\supit{f}, M.~Levi\supit{a},
 D.~Levin\supit{b}, E.~Linder\supit{a},
S.~Loken\supit{a},
R.~Malina\supit{m}, R.~Massey\supit{n},
T.~McKay\supit{b}, S.~McKee\supit{b},
R.~Miquel\supit{a},
E.~M\"{o}rtsell\supit{c}, N.~Mostek\supit{h}, S.~Mufson\supit{h}, J.~Musser\supit{h},
P.~Nugent\supit{a}, H.~Oluseyi\supit{a}, R.~Pain\supit{d}, N.~Palaio\supit{a},
D.~Pankow\supit{f}, S.~Perlmutter\supit{a}, R.~Pratt\supit{f},
E.~Prieto\supit{m},
A.~Refregier\supit{n},
 J.~Rhodes\supit{o},
K.~Robinson\supit{a}, N.~Roe\supit{a}, M.~Sholl\supit{f}, M.~Schubnell\supit{b},
G.~Smadja\supit{p}, G.~Smoot\supit{f}, A.~Spadafora\supit{a},
G.~Tarl\'e\supit{b}, A.~Tomasch\supit{b}, H.~von der Lippe\supit{a},
D.~Vincent\supit{d},
J-P.~Walder\supit{a}, G.~Wang\supit{a}}
\skiplinehalf
\supit{a}Lawrence Berkeley National Laboratory, Berkeley CA, USA\\
\supit{b}University of Michigan, Ann Arbor MI, USA\\
\supit{c}University of Stockholm, Stockholm, Sweden\\
\supit{d}CNRS/IN2P3/LPNHE, Paris, France\\
\supit{e}University of Pennsylvania, Philadelphia PA, USA\\
\supit{f}University of California, Berkeley CA, USA\\
\supit{g}CNRS/IN2P3/CPPM, Marseille, France\\
\supit{h}Indiana University, Bloomington IN, USA\\
\supit{i}American Astronomical Society, Washington DC, USA\\
\supit{j}California Institute of Technology, Pasedena CA, USA\\
\supit{k}Space Telescope Science Institute, Baltimore MD, USA\\
\supit{l}Case Western Reserve University, Cleveland OH, USA\\
\supit{m}CNRS/INSU/LAM, Marseille, France\\
\supit{n}Cambridge University, Cambridge, UK\\
\supit{o}Goddard Space Flight Center, Greenbelt MD, USA\\
\supit{p}CNRS/IN2P3/IPNL, Lyon, France
}

\authorinfo{Send correspondence to A. Kim: E-mail: agkim@lbl.gov}

\begin{document} 
  \maketitle 

\begin{abstract}
The Supernova / Acceleration Probe (SNAP) is a proposed
space-borne observatory that will survey the sky with a wide-field
optical/near-infrared (NIR)  imager.  The images produced by SNAP will have
an unprecedented combination of
depth, solid-angle, angular resolution, and temporal sampling.
For 16 months each,
two 7.5 square-degree fields will be observed every four days
to a magnitude depth of $AB=27.7$ in each of
the SNAP filters, spanning 3500--17000\AA.
Co-adding images over all epochs will give
$AB=30.3$ per filter.  In addition, a 300 square-degree
field will be surveyed to $AB=28$ per filter, with no repeated
temporal sampling.
Although the survey strategy
is tailored for supernova and weak gravitational lensing observations,
the resulting data will support a broad range of auxiliary science
programs.
\end{abstract}



\section{INTRODUCTION}
\label{sect:intro}  

The unexpected discovery that the Universe's
expansion is accelerating, as measured by supernova
experiments
\cite{42SNe_98,riess_acc_98}
and independently confirmed by cosmic-microwave-background
experiments\cite{Balbi:2000,Lange:2001}, 
implies 
that some heretofore unknown form of energy is driving
the Universe's dynamics.
The existence of this so-called ``dark energy'' lies beyond
the current framework of elementary particles and thus has
profound implications for
fundamental physics.
The challenge now is to measure the physical properties of
this dark energy;
the most well-developed approach
is with a next generation high-redshift supernova search and
discovery experiment.

The Supernova / Acceleration Probe (SNAP)
is proposed in this spirit.  As a dedicated space-borne observatory,
SNAP will provide supernova data, in the form of light curves
and spectra, of unprecedented quality.  The photometric instrumentation
suite is tailored specifically for the needs of the supernova program;
a wide-field imager in the optical and NIR provides a high
discovery rate of $z<1.7$ Type Ia supernovae
(SNe Ia) and allows for multiplexed followup observations, $\sim 50$ supernovae
within a single exposure.
The supernova fields will be revisited every four days for 16 months, 
  providing light curves for at least several months in the rest frame
  of each supernova.  The optical light of the supernova
will observed using filter set spanning 3500--17000\AA\ 
in the observer frame.

Also part of the SNAP primary mission
is a weak gravitational lensing survey.  Lensing provides
an independent and complementary
measurement of the cosmological parameters through
the mapping of galaxy shape distortions induced by
mass inhomogeneities in the Universe.
The strengths that make SNAP excellent for supernova observations
apply to lensing as well; a wide-field imager in space with
stable and narrow point-spread-functions can provide large
survey areas, accurate shape measurements, and high galaxy angular-surface
densities.
The SNAP supernova
fields will serve as a deep lensing field while a second
larger-solid-angle field specifically tailored for lensing
will be observed to a shallower depth.

In this paper, we quantify the expected
depth, solid-angle, and
time resolution of the SNAP SNe and weak lensing surveys.
In \S\ref{mission:sec} we describe the SNAP mission:
 the important properties of the
telescope and camera as relevant to imaging
and the observing cadence and exposure times
of the primary SNAP science missions.
The depths of the surveys naturally
produced by these programs are given
in \S\ref{depth:sec}.  A sampling of possible auxiliary science that
can be done with these data is given in \S\ref{science:sec}
and a brief summary is given in \S\ref{summary:sec}.

\section{DESCRIPTION OF MISSION}
\label{mission:sec}
In this section, we describe the SNAP telescope and its
instrumentation suite as relevant to its imaging capabilities.
The observing program for the primary supernova and lensing
missions are also detailed.  It is based on these properties
that we calculate
the depth of the resulting surveys.
The numbers provided here and throughout this paper are
SNAP specifications that are subject to change as we work on
the conceptual design.
\subsection{Telescope and Instrumentation}
The important properties of the SNAP telescope\cite{lampton:2002} are given
in Table~\ref{telescope:tab}.  SNAP has a 2-m primary
aperture and 16\% obscuration from the full baffling.
The spot sizes from ray tracing of the telescope optics are less than
0.05'' RMS over the focal plane.  The optical telescope assembly
is composed of four silver-coated reflectors each with 98\% throughput.
\begin{table}[h]
\caption{Parameters of the SNAP telescope that
are relevant
to determining imaging capabilities.} 
\label{telescope:tab}
\begin{center}       
\begin{tabular}{|c|c|c|c|c|c|c|} 
\hline
Telescope & Primary & Secondary &
Primary & Spot Size & Throughput & Jitter\\
& Aperture (m) & Aperture (m) & Obscuration & (arcsec) &(@ 1$\mu$m) & (arcsec)\\
\hline 
SNAP & 2.0 & 0.4 & 0.16 & 0.05 &0.92 & 0.02\\ \hline
\end{tabular}
\end{center}
\end{table}

The SNAP camera\cite{bebek:2002} has
$f/\#=10.83$ and throughput of 70\%.
The camera tiles 0.7 square degrees, split in area between
LBNL CCD's\cite{Groom-nim} and 1.7$\mu$m cutoff HgCdTe devices\cite{tarle:2002}.
The detector
properties are given in Table~\ref{det:tab}.  The detector pixel sizes
give undersampled images; spatial resolution will be recovered
by taking several dithered images for each pointing\cite{bernstein:etc}.
When imaging,
the detector
noise is sub-dominant to the zodiacal background
and thus has a negligible effect on the error budget.
\begin{table}[h]
\caption{Parameters of the SNAP
detectors that are relevant to determining imaging capabilities.} 
\label{det:tab}
\begin{center}       
\begin{tabular}{|c|c|c|c|c|c|} 
\hline
Detector & Pitch & Read Noise & Dark Current & Diffusion & Peak QE\\
&($\mu$m)& (e$^-$/pixel)& (e$^-$/sec/pixel) & ($\mu$m)&\\
\hline 
LBNL CCD\cite{Groom-nim} & 10.5 & 4 & 0.002 & 4 & 0.92\\ \hline
HgCdTe\cite{tarle:2002} & 18 & 5 & 0.02 & 5 & 0.6\\ \hline
\end{tabular}
\end{center}
\end{table}

\begin{figure}[b]
   \begin{center}
   \begin{tabular}{c}
   \includegraphics[height=7cm]{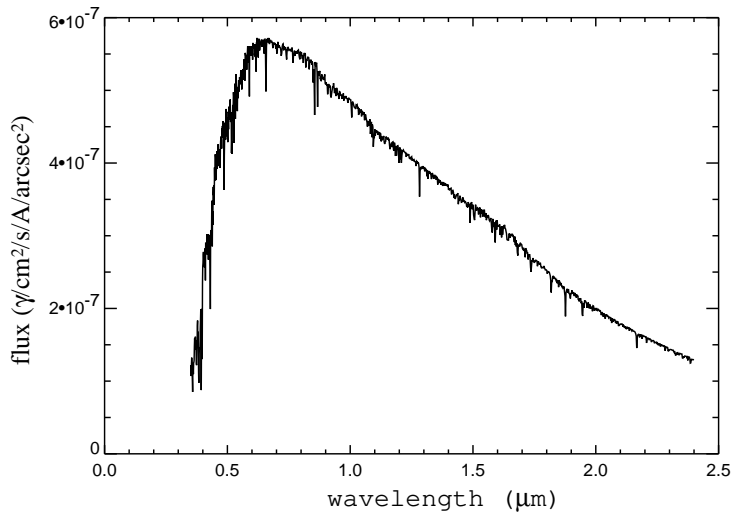}
   \end{tabular}
   \end{center}
   \caption[example] 
   { \label{zodi:fig} 
The zodiacal background will be the dominant source of background
light for SNAP.  Shown is the zodiacal photon flux toward the north
ecliptic pole, the planned location for one of the SNAP supernova
surveys.}
   \end{figure}

The zodiacal background will be the dominant source of
background light given SNAP's orbit and shielding of Earth-shine.  The
relatively faint wavelength-dependent background toward the
ecliptic poles is
shown in Figure~\ref{zodi:fig}\cite{zodi02}.  The cosmic-ray flux of
$2\times 10^{-4}$/sec/pix
will make multiple measures necessary to avoid
significant contamination of the images, as will be discussed
in \S\ref{depth:sec}.

The PSF is diffraction dominated but we include effects such as
CCD diffusion, telescope jitter, and the telescope spot blur.
At bluer wavelengths, the contribution of these secondary blurs do have
an important effect on the PSF.  For our photometric measurements,
it is the PSF size that
determines the noise contribution
of sky background.

 The SNAP filter set, their
shapes, number, and distribution in wavelength space, is currently
under review.
The set that we consider in this note
consists of nine
Johnson B filters logarithmically distributed in wavelength
with effective wavelengths at $4400 \times 1.15^n$\AA\ 
for $n \in \{0,1,...,8\}$. 
  We have a fixed filter design;
the six optical filters are uniformly distributed
over the CCD's while the three NIR filters occupy equal areas
of the HgCdTe devices.  An individual optical filter tile subtends
$2.9' \times 2.9'$ and an individual NIR-filter tile
subtends $5.8' \times 5.8'$.
Summing over the tiles, each
optical filter covers 0.056 square degrees over 24 squares
while each NIR
filter covers 0.112 square degrees over 12 squares to
constitute the entire 0.7 square degree field.

\begin{figure}[tb]
   \begin{center}
   \begin{tabular}{c}
   \includegraphics[height=10cm]{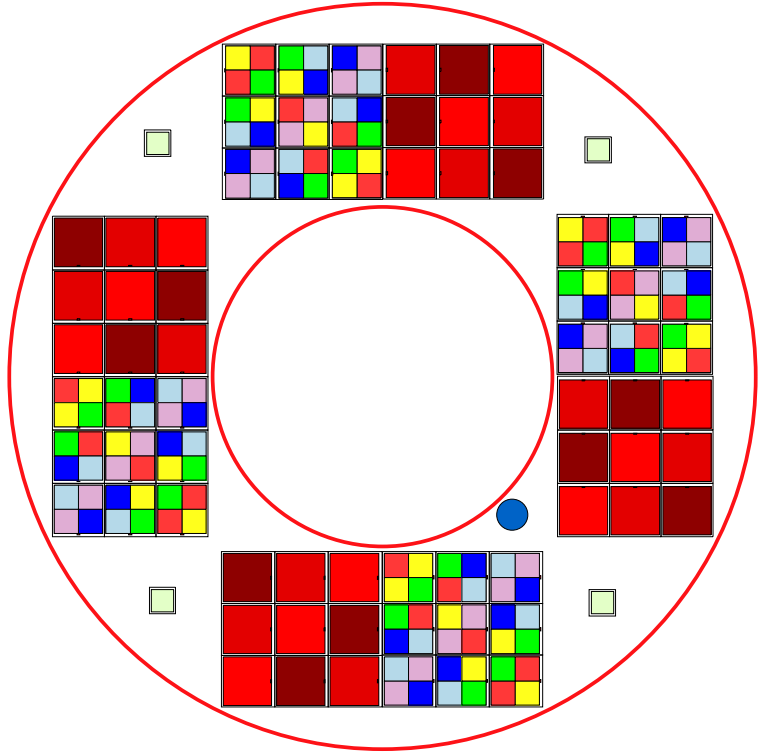}
   \end{tabular}
   \end{center}
   \caption[example] 
   { \label{imager:fig} 
The layout for the SNAP imager.  Detectors tile 0.7 square
degrees of the focal plane.  In the fixed filter scheme,
six optical filters are mated to CCD's and three NIR filters
are mated to HgCdTe detectors.  An individual optical-filter tile
subtends $2.9' \times 2.9'$ and an individual NIR-filter tile
subtends $5.8' \times 5.8'$.  The four star guiders are shown as
isolated squares located between
the science detectors.  The circle at the inner part of the annulus
is the light-access port for the spectrograph.}
   \end{figure}

The detectors are arrayed in an annulus; the SNAP
three-mirror-anastigmatic design has a flat pickoff
mirror near the Cassegrain focus that totally vignettes
the central region of the field.
The 90$^\circ$ rotational
symmetry of the filter layout allows a fixed side of SNAP
to always face the sun while maintaining a consistent
footprint of the fields over an entire year. 
Figure~\ref{imager:fig} shows the layout of the imager in the focal
plane, and the relative sizes and positions of the filters.

\subsection{Observing Strategy}
The two primary SNAP science programs have individually-designed observing
schedules which are described here and summarized in Table~\ref{obs:tab}.
\subsubsection{Supernova Program}
The supernova survey is comprised of two halves, both in time and
in position. The initial survey will cover
7.5 square degrees over sixteen months pointing toward a
field near the north ecliptic
pole.  After an intervening weak-lensing survey, a similar field toward
the south ecliptic pole will be observed. 

Observations in multiple bands allow SN-frame $B$ and $V$
coverage for SNe from $0<z<1.7$.
As shown in Figure~\ref{imager:fig},
the filters are arrayed in a checker-board pattern.  In the absence of
a filter
wheel, multi-band exposures of a point of sky will be achieved by
shift-and-stare observations where the
pointings are shifted by the 2.9'  width of the optical filters.
 Each pointing
will consist of four 300-second exposures; the multiple exposures
are for cosmic-ray rejection and dithering of our undersampled pixels.
Within a scan, over one hundred
pointings are required to cover the 7.5 square degrees
in all filters to the desired depth.
The wider NIR filters will experience twice
the exposure time of the optical filters.
A scan of the north (south) field will
be repeated every four days for 16 months for a total 120 scans.

Forty percent of the SNAP mission will be spent doing targeted spectroscopy
of supernovae.
Imaging occurs simultaneously during these spectroscopic observations.  The
resulting images will cover random positions and orientations within the SNAP
field and can be used to increase the depth of the survey.  However,
these extra images
are not considered in the calculations given in this paper.

\subsubsection{Weak Lensing Program}
The weak lensing program calls for a five-month wide-field survey
to obtain as much solid-angle as possible within
the constraints of telemetry.  Repeat observations are unnecessary.
Multi-filter data, particularly in the NIR, is desired for accurate
photometric-redshift determination.
The specific geometry of this field is yet to be determined.
The important observing parameters of this and the
supernova programs are summarized in Table~\ref{obs:tab}.

\begin{table}[h]
\caption{The SNAP Surveys.} 
\label{obs:tab}
\begin{center}       
\begin{tabular}{|c|c|c|c|c|c|}
\hline
Program & Solid Angle & Exposure & Cadence & Visits\\
&per filter (sq.\ deg.)& per scan (s)& (days)& \\
\hline 
Supernova & $7.5 \times 2$ & Optical: $4\times 300$ & 4 & 120\\
 & &NIR: $8 \times 300$ & &\\ \hline
Lensing & 300 & Optical: $4\times 500$ & \ldots & 1 \\
 & &NIR: $8 \times 500$ & &\\ \hline
\end{tabular}
\end{center}
\end{table}

\section{DEPTH OF OBSERVATIONS}
\label{depth:sec}
The telescope and camera properties of SNAP have been modeled and
incorporated into
an
advanced exposure-time calculator (ETC)\cite{bernstein:etc}.
Besides having all the bells and whistles of a standard ETC,
our ETC includes unique handling of the pixel
response function, undersampling, dithering, and probabilistic
cosmic-ray rejection.
As mentioned in \S\ref{mission:sec}, SNAP will rely on dithering
to recover spatial resolution from its undersampled pixels.

The high cosmic-ray flux produces a non-trivial reduction of
effective exposure times; pixels from a single exposure
that are contaminated by a cosmic
ray are assumed to be recognized through median filtering
and dropped in the dithered reconstruction.
Short individual exposure times limit the contamination: 300 second
exposures give a 68\% probability that
there will be no cosmic-ray contamination at a given position
on any of the four dithers
that make up a pointing.

The magnitude depths for individual scans and co-added images of
the SNAP supernova fields are calculated for each filter.  The limiting
magnitude for any given point is probabilistic, due to the
random occurrence of cosmic rays.  Table~\ref{depth:tab}
shows the 50th-percentile limiting AB magnitude for a $S/N=5$ point source
for each filter in the surveys.
\begin{table}[h]
\caption{The SNAP 50th-percentile
AB magnitude survey depth for a point source $S/N=5$.
Random cosmic-ray hits make the $S/N$ for a given position probabilistic.
The choice of filter set is currently subject to optimization studies;
the filters and depths presented here are meant to be illustrative.} 
\label{depth:tab}
\begin{center}       
\begin{tabular}{|c|c|c|c|c|c|c|}
\hline
Filter & $\lambda_{eff}$(\AA) & $\Delta \lambda$(\AA)& \multicolumn{2}{c|}{SN Survey (AB mag)} & Lensing Survey\\ \cline{4-5}
& & &Scan & Co-added Scans& (AB mag) \\ \hline
1 & 4400& 1000 & 27.9	&30.6	&28.3\\ \hline 
2 & 5060& 1150  &27.8	&30.5	&28.2\\ \hline
3 & 5819& 1323 & 27.8	&30.4	&28.1\\ \hline
4 & 6692& 1521 & 27.7	&30.4	&28.1\\ \hline 
5 & 7696& 1749 & 27.7	&30.3	&28.0 \\ \hline
6 & 8850 & 2011 & 27.5	&30.2	&27.9\\ \hline
7 & 10178& 2313 & 27.5	&30.2	&27.8\\ \hline 
8 & 11704& 2660 & 27.4	&30.1	&27.8\\ \hline
9 & 13460 &3059 & 27.4	&30.0	&27.7 \\ \hline
\end{tabular}
\end{center}
\end{table}

The SNAP observing strategy provides remarkably even depth over
the range of filters.  For a given filter,
individual scans of the supernova and lensing surveys are only $\sim 0.75$
magnitudes shallower than the Hubble Deep Fields (HDFs) while the SN fields
co-added over time are
$\sim 1.5$ magnitudes deeper than the HDFs\cite{williams_96}.
SNAP has the additional advantage of
having nine filters observing to this depth, compared
to the four filters of the HDFs, and 9000 times the area;
when these data are combined,
the limiting magnitude increases by 0.6 magnitudes.

SNAP fields will contain many faint diffuse galaxies whose detection
is important for the weak-lensing survey, and for other potential science
projects.
The limiting magnitudes for Gaussian-aperture photometry of an exponential-disk
galaxy with FWHM=0.12'' are shown in Table~\ref{depth_diff:tab}.
\begin{table}[h]
\caption{The SNAP AB magnitude survey depth for an exponential-disk galaxy
with FWHM=0.12'' with $S/N=10$.
The choice of filter set is currently subject to optimization studies;
the filters and depths presented here are meant to be illustrative.} 
\label{depth_diff:tab}
\begin{center}       
\begin{tabular}{|c|c|c|c|c|c|c|}
\hline
Filter & $\lambda_{eff}$(\AA) & $\Delta \lambda$(\AA)& \multicolumn{2}{c|}{SN Survey (AB mag)} & Lensing Survey\\ \cline{4-5}
& & &Scan & Co-added Scans& (AB mag) \\ \hline
1 & 4400& 1000& 26.4	&29.1	&26.8 \\ \hline 
2 & 5060& 1150 &26.3	&29.0	&26.7 \\ \hline
3 & 5819& 1323 &26.3	&29.0	&26.6\\ \hline
4 & 6692&1521 & 26.2	&28.9	&26.6\\ \hline 
5 & 7696& 1749 & 26.3	&28.9	&26.6\\ \hline
6 & 8850&  2011 & 26.2	&28.8	&26.5\\ \hline
7 & 10178& 2313 & 26.3 &28.9	&26.6\\ \hline 
8 & 11704&  2660 &26.2	&28.9	&26.6\\ \hline
9 & 13460& 3059 & 26.2	&28.9	&26.5\\ \hline
\end{tabular}
\end{center}
\end{table}

\section{SCIENCE}
\label{science:sec}
In this section we give a brief discussion of possible science that
can be obtained from the SNAP surveys.  This list
is by no means complete in its breadth nor depth.
The expected results from
the primary SNAP science missions are discussed in a
companion paper\cite{aldering:2002}.

The Sloan Digital Sky Survey(SDSS)\cite{York:2000} and
HDFs \cite{williams_96,williams:2000} have demonstrated
the vast range of
science that can be obtained from wide and deep multi-band surveys.
SNAP will produce surveys
that dwarf the 0.0016 square degrees size of the HDFs
and go even deeper with
time-sampling
for its supernova fields.
The SNAP lensing field is about the same size as the Sloan Southern
Survey and CFHT Legacy Survey fields but several magnitudes deeper. 
This combination of depth, temporal coverage, filter coverage over
a broad wavelength range, diffraction-limited seeing,
and wide field make SNAP imaging surveys uniquely powerful in the study
of a wide range of objects:
\begin{itemize}
\item Galaxies --- SNAP fields
will contain $> 5\times 10^7$ galaxies within detection threshold.
Statistical studies
are possible with such a large sample, e.g.\
the determination of the galaxy luminosity function and color
distributions as a function of redshift.
The depth allows discovery
of low-surface-brightness and very high-redshift
galaxies.  Accurate photometric redshifts and
information on galaxy evolution will be available from
the multi-band and in particular the NIR photometry.
High-resolution images will provide
a view of the internal structure of galaxies and their kinematic
interactions with each other.
\item Galaxy clusters --- Wide-field imaging and photometric redshifts
allow easy identification of galaxy clusters.  The epoch
of galaxy-cluster formation is tightly linked with the mass
density of the Universe, $\Omega_M$, providing an independent
cosmological measurement complementary to SNAP's primary missions.
\item Quasars --- The NIR photometry extends the redshift
range for quasar
discovery ($6.3 <z < 12$) using colors and
dropout surveys.  Discoveries will also move much fainter into the
quasar luminosity function.  The highest redshift quasars can be used to
map the reionization history of the Universe through the Gunn-Peterson
effect. SNAP's ability to identify diffuse objects associated with quasars
may present many opportunities
for the study of galaxy formation.
\item Transients/Variables --- The discovery and observation
of SNe Ia are the primary goals of SNAP, but transient ``backgrounds''
are interesting in their own right: quasars, active-galactic-nuclei,
gamma-ray-burst
optical counterparts, supernovae of other types, variable stars,
and eclipsing binaries.  Of particular interest to cosmology is time-delay
studies with the
expected large number of strongly lensed variables.
Gravitational microlensing surveys of stars and quasars to measure
dark matter are also
possible.
\item Stars --- Faint limiting magnitudes and excellent star-galaxy
separation will yield faint dwarf and halo stars.  Proper motion
can be detected with high-resolution and a long time baseline.
The geometry and substructure of the halo and disk in the direction
of the SNAP fields can be mapped.
\item Solar-system objects ---  The peculiar motion in the time-series
data will facilitate the identification of local objects such as asteroids
and Kuiper-belt objects.
\end{itemize}

The output from the SDSS has demonstrated how
the natural byproducts of a wide-field survey can produce
scientific yield well beyond the
scope of its primary purpose.
Individual objects found on SDSS images are routinely observed
spectroscopically at the largest telescopes in the world,
fulfilling the
historical trend of small-aperture telescope imaging feeding
targets
for large-aperture telescope spectroscopy.  
The SNAP surveys will provide a similar opportunity in working
with NGST and the next generation of ground-based
wide-aperture telescopes.

\section{SUMMARY}
\label{summary:sec}
As we have discussed, SNAP will provide a combination of depth,
solid-angle, angular resolution, and temporal sampling
heretofore unachieved.
The primary science missions of SNAP will provide survey
fields in
filters spanning 3500--17000\AA.
\begin{itemize}
\item A 300 square-degree field to AB mag $\sim 28$ at $S/N=5$
in each filter.
\item Two 7.5 square-degree fields observed every four
days 120 times in all filters.
Each observation reaches AB mag $\sim 27.7$ at $S/N=5$
in each filter.
\item The co-added sum of all visits from the preceding survey:
two 7.5 square-degree fields to AB mag $\sim 30.3$ at $S/N=5$
in each filter.
\end{itemize}

We have restricted our discussion to surveys produced by the
principal science missions of SNAP\footnote{The numbers presented here describe the SNAP reference mission.
They are subject to change as trade studies are performed to
optimize the design of the experiment.
The reader is advised to exercise caution in extrapolating
these results to other possible survey missions;
the SNAP primary-mission fields have continuous visibility, allow for stable
satellite orientation, ensure shielding of background light, and
point to regions of relatively low zodiacal background.}.
Additional imaging surveys will be produced from data taken
after completion of
the SNAP primary missions;
a guest observing program
is envisioned
filling the remaining satellite lifetime to allow the full potential
of SNAP to be realized.
\acknowledgments     
This work was supported by the Director, Office of Science, of the
U.S.   Department of Energy under Contract No. DE-AC03-76SF00098.


\end{document}